\documentclass[aps,prd,twocolumn,nofootinbib,superscriptaddress]{revtex4-2}

\usepackage{amsmath,amssymb}
\usepackage{graphicx}
\usepackage{bm}
\usepackage[hidelinks]{hyperref}

\newcommand{\Ob}{\Omega_b}
\newcommand{\Oc}{\Omega_c}
\newcommand{\Om}{\Omega}
\newcommand{\dd}{\mathrm{d}}

\begin{document}

\title{Phase-space structure of cosmological thermogravity}

\author{Charlotte Brereton}
\author{Jo\~ao Magueijo}
\author{Mar Alí Caballo}
\email{magueijo@ic.ac.uk}
\affiliation{Abdus Salam Centre for Theoretical Physics, Imperial College London, Prince Consort Road, London SW7 2BZ, United Kingdom}

\begin{abstract}
An emergent gravity model was shown to contain an associated emergent breaking of Lorentz invariance and violations of stress-energy conservation. Its simplest flat dust solution accelerates at late times without vacuum energy and has already been used as an analytic benchmark for observational tests.  We study the wider homogeneous dynamics and solutions of this theory.  For a single non-conserved dust fluid, the variables $\Om=\rho/(3H^2)$ and $u=(HL_2)^{-1}$ form a two-dimensional autonomous system with an exact first integral.  The complete expanding solution space separates into three classes: trajectories ending on an asymptotically de Sitter fixed line, a critical trajectory corresponding to the analytic solution of Ref.~\cite{IsicheiMagueijo2026}, and trajectories that turn around and recollapse.  We then consider the phenomenologically cleaner model in which only cold dark matter violates energy conservation (while the baryons and photons behave as normal).  The phase space becomes three-dimensional, with a line of matter repellers and a line of vacuum fixed points.  Each normally attracting vacuum point with $0<u_0<3$ possesses a two-dimensional stable sheet.  The limit in which the CDM abundance tends to zero is singular: finite-$u_0$ sheets meet the exactly baryonic plane on trajectories containing a cosmological-constant-like integration constant, whereas the zero-integration-constant baryonic solution remains at $\Ob=1$ and runs to $u\to\infty$.  The critical endpoint $u_0=3$ separates these late-time behaviours and generalizes the special single-fluid solution.
\end{abstract}

\maketitle

\section{Introduction}

The thermodynamic interpretation of gravity has its roots in black-hole entropy and temperature~\cite{Bekenstein1973,Hawking1975} and Jacobson's derivation of the Einstein equation as an equation of state~\cite{Jacobson1995}.  The original idea has since been extended to non-equilibrium settings and higher-curvature theories~\cite{Eling2006,Paranjape2006,ChircoLiberati2010,Padmanabhan2010}, to entanglement-equilibrium formulations~\cite{Jacobson2016}, and to other emergent-gravity constructions~\cite{Verlinde2011,Verlinde2017}. Much of this leads to previously known theories of gravity, thereby shedding new light on them. 

Ref.~\cite{IsicheiMagueijo2026} takes a different approach: could we look at the thermodynamic emergent route as conceptually primary, and see whether it guides us to emergent theories which are truly new, previously unknown macroscopic constructions? Noting that the usual thermodynamic argument is based on pure heat exchange, we discover that allowing a non-degenerate thermodynamic cycle gives, at the macroscopic level, trace-free Einstein equations supplemented by a preferred-frame violation of stress-energy conservation.  Because the metric equation is trace-free, the construction is closely related to the long history of unimodular and trace-free gravity~\cite{AndersonFinkelstein1971,Unruh1989,HenneauxTeitelboim1989,NgVanDam1991,Smolin2009,PadillaSaltas2015}; unlike the conserved case, however, the additional non-conservation law obstructs the usual reinstatement of a cosmological constant as a free integration constant.  This gives a new angle on the cosmological-constant problem~\cite{Weinberg1989,Padilla2015} and, for homogeneous matter (including pure dust), allows matter production to drive late-time acceleration without introducing gravitating vacuum energy.  The preferred-frame aspect also places the theory within the broader phenomenology of Lorentz-symmetry breaking~\cite{Liberati2013}.

The cosmological solution emphasized in Ref.~\cite{IsicheiMagueijo2026} is particularly simple,
\begin{equation}
 a(t)=a_\star t^{2/3}\exp\left(\frac{t}{3L_2}\right),
 \label{eq:prlsolintro}
\end{equation}
and has subsequently been used as the analytic benchmark of the universal non-conservation model in the first background-level observational analysis of the theory~\cite{LeeEtAl2026}.  The same work also studied a non-universal implementation in which only cold dark matter (CDM) violates energy conservation, while baryons and photons retain their standard evolution.  That model is observationally cleaner because the usual late-time rulers and clocks are not directly modified at background level~\cite{LeeEtAl2026}.

Here we ask a complementary question: what is the global phase-space structure of these cosmologies, and how special is solution Eq.~\eqref{eq:prlsolintro}?  Dynamical-system methods are well suited to classifying cosmological asymptotics~\cite{Bahamonde2018}. Using these techniques, we first show that the universal dust model is exactly integrable in phase space and that Eq.~\eqref{eq:prlsolintro} is a critical separatrix between de Sitter-like expansion and recollapse.  We then extend the analysis to conserved baryons plus non-conserved CDM.  This reveals a family of stable sheets ending on a line of vacuum fixed points and, importantly, a non-uniform $\Oc\rightarrow0$ limit tied to the cosmological-constant-like integration constant familiar from trace-free gravity~\cite{PadillaSaltas2015}.

Throughout this paper we will set the spatial curvature to zero ($K=0$), use units such that $8\pi G=1$, assume the new parameter in the emergent theory is positive, $L_2>0$, and use the number of e-foldings $N\equiv\ln a$ as time variable.

\section{Universal dust model}
\label{sec:universal}

For one pressureless component the background equations of Ref.~\cite{IsicheiMagueijo2026} reduce to
\begin{equation}
 \dot H=-\frac{\rho}{2},
 \qquad
 \dot\rho+3H\rho=\frac{\rho}{L_2}.
 \label{eq:bg}
\end{equation}
Defining
\begin{equation}
 \Om\equiv\frac{\rho}{3H^2},
 \qquad
 u\equiv\frac{1}{HL_2},
 \qquad
 {}'\equiv\frac{\dd}{\dd N},
 \label{eq:variables}
\end{equation}
we obtain the autonomous system
\begin{align}
 \Om'&=\Om\left[u+3(\Om-1)\right],
 \label{eq:Omeq}\\
 u'&=\frac32\Om u.
 \label{eq:ueq}
\end{align}
The point $(\Om,u)=(1,0)$ has eigenvalues $(3,3/2)$ and is therefore a past attractor (future repeller).  The entire line $\Om=0$ consists of fixed points.  At $(0,u_\star)$ the transverse eigenvalue is $u_\star-3$; hence the segment $0<u_\star<3$ is normally attracting, while $u_\star=3$ is non-hyperbolic.

\begin{figure}[t]
\centering
\includegraphics[width=\columnwidth]{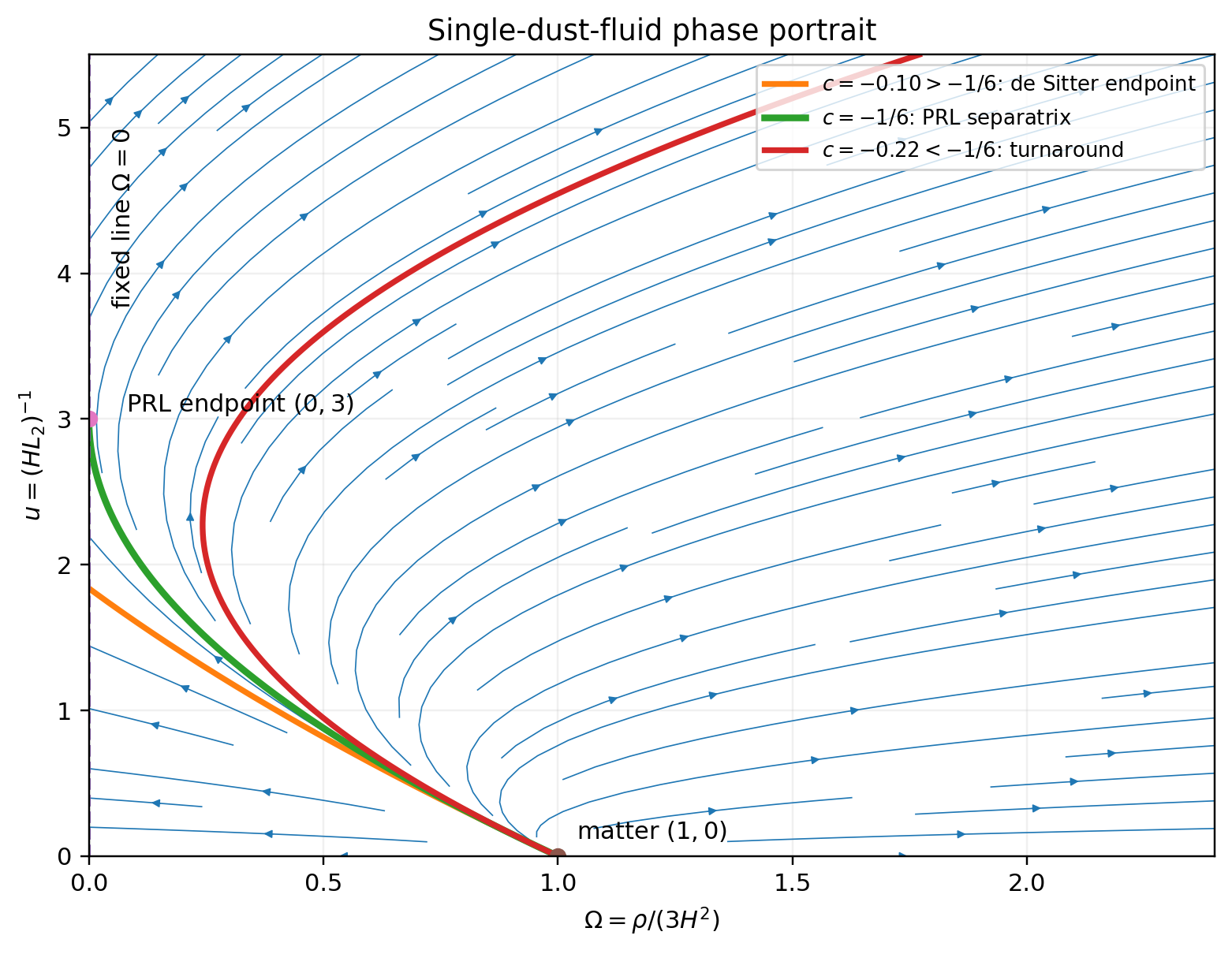}
\caption{Phase portrait of the universal dust system, Eqs.~\eqref{eq:Omeq}--\eqref{eq:ueq}.  The critical $c=-1/6$ trajectory is the solution used in Ref.~\cite{IsicheiMagueijo2026}.  Trajectories with $c>-1/6$ terminate on the attracting part of the fixed line $\Om=0$; those with $c<-1/6$ reach $H=0$, where $(\Om,u)$ become singular although the physical variables remain finite.}
\label{fig:universal}
\end{figure}

\subsection{First integral and exact classification}

The phase trajectories are analytic.  Dividing Eq.~\eqref{eq:Omeq} by Eq.~\eqref{eq:ueq} gives
\begin{equation}
 \frac{\dd\Om}{\dd u}-\frac{2}{u}\Om
 =\frac23-\frac{2}{u},
\end{equation}
so that
\begin{equation}
 c\equiv
 \frac{\frac32(1-\Om)-u}{u^2}
 =\text{const},
 \quad
 \Om=1-\frac23u-\frac23c\,u^2.
 \label{eq:firstintegral}
\end{equation}
The same constant follows directly from Eq.~\eqref{eq:bg}:
\begin{equation}
 \dot H=\frac{H}{L_2}-\frac32H^2+C,
 \qquad c\equiv C L_2^2.
 \label{eq:Hfirst}
\end{equation}
Thus a single number labels every orbit in Fig.~\ref{fig:universal}.

It is useful to display the corresponding exact spacetime solutions.  Let
\begin{equation}
 \Delta\equiv1+6c.
\end{equation}
For $\Delta>0$, defining $s\equiv\sqrt{\Delta}$ and $\tau=t-t_B$, the big-bang branch is
\begin{align}
 H&=\frac{1}{3L_2}
 \left[1+s\coth\left(\frac{s\tau}{2L_2}\right)\right],\\
 a&=a_\star e^{\tau/(3L_2)}
 \sinh^{2/3}\left(\frac{s\tau}{2L_2}\right),\\
 \rho&=\frac{s^2}{3L_2^2}
 \operatorname{csch}^2\left(\frac{s\tau}{2L_2}\right).
 \label{eq:dSbranch}
\end{align}
It approaches
\begin{equation}
 H_\star=\frac{1+s}{3L_2},
 \qquad
 u_\star=\frac{3}{1+s}<3,
 \label{eq:ustar}
\end{equation}
with $\rho\to0$.  Thus the geometry becomes asymptotically de Sitter even though the material density vanishes.  In phase space,
\begin{equation}
 \Om\propto e^{-(3-u_\star)N},
 \qquad
 u_\star-u\simeq
 \frac{3u_\star}{2(3-u_\star)}\Om.
 \label{eq:genericasymptotic}
\end{equation}

At the critical value $c=-1/6$, $\Delta=0$ and Eq.~\eqref{eq:dSbranch} reduces to
\begin{align}
 H&=\frac{1}{3L_2}+\frac{2}{3\tau},\\
 a&=a_\star \tau^{2/3}e^{\tau/(3L_2)},
 \qquad
 \rho=\frac{4}{3\tau^2},
 \label{eq:criticalsolution}
\end{align}
which is precisely the solution of Ref.~\cite{IsicheiMagueijo2026}.  Its phase trajectory is
\begin{equation}
 \Om=\left(1-\frac{u}{3}\right)^2,
\end{equation}
and it joins $(1,0)$ to the non-hyperbolic endpoint $(0,3)$.  The approach is algebraic,
\begin{equation}
 \Om\sim\frac{4}{9N^2},
 \qquad
 3-u\sim\frac{2}{N},
 \label{eq:criticalasymptotic}
\end{equation}
rather than exponential.  The commonly used analytic cosmology is therefore a separatrix, not a generic future attractor.

Finally, for $\Delta<0$ define $q\equiv\sqrt{-\Delta}$.  Then
\begin{align}
 H&=\frac{1}{3L_2}
 \left[1+q\cot\left(\frac{q\tau}{2L_2}\right)\right],\\
 a&=a_\star e^{\tau/(3L_2)}
 \sin^{2/3}\left(\frac{q\tau}{2L_2}\right),\\
 \rho&=\frac{q^2}{3L_2^2}
 \csc^2\left(\frac{q\tau}{2L_2}\right).
 \label{eq:recollapse}
\end{align}
The expanding solution reaches $H=0$ at finite time with finite $a$ and $\rho$.  Since $\dot H=-\rho/2<0$, it necessarily crosses into contraction and ends in a finite-time crunch.  The divergence $u,\Om\to\infty$ in Fig.~\ref{fig:universal} is therefore only the coordinate singularity associated with $H\to0$.

The three possible futures are consequently separated exactly by the critical value $c=-1/6$:
\begin{itemize}
    \item $c>-1/6$: Asymptotic de Sitter expansion. 
    \item $c=-1/6$: Critical thermogravity solution in \cite{IsicheiMagueijo2026}. 
    \item $c<-1/6$: Turnaround and recollapse.
\end{itemize}

\section{Conserved baryons and non-conserved CDM}
\label{sec:cdm}

The universal model is useful analytically, but a cleaner phenomenological implementation leaves baryons conserved and confines the violation to CDM~\cite{IsicheiMagueijo2026,LeeEtAl2026}.  Neglecting radiation at late times,
\begin{align}
 \dot\rho_b+3H\rho_b&=0,\\
 \dot\rho_c+3H\rho_c&=\frac{\rho_c}{L_2},\\
 \dot H&=-\frac12(\rho_b+\rho_c).
 \label{eq:twocomponentbackground}
\end{align}
Introduce
\begin{equation}
 \Ob=\frac{\rho_b}{3H^2},
 \qquad
 \Oc=\frac{\rho_c}{3H^2},
 \qquad
 u=\frac{1}{HL_2}.
\end{equation}
The autonomous system is
\begin{align}
 \Ob'&=3\Ob(\Ob+\Oc-1),
 \label{eq:Ob}\\
 \Oc'&=\Oc\left[u+3(\Ob+\Oc-1)\right],
 \label{eq:Oc}\\
 u'&=\frac32(\Ob+\Oc)u.
 \label{eq:u3d}
\end{align}
There are again two continuous sets of fixed points.  The matter line
\begin{equation}
 M(\beta)=(\beta,1-\beta,0),
 \qquad 0\leq\beta\leq1,
 \label{eq:matterline}
\end{equation}
has eigenvalues $(3,3/2,0)$ and is a past attractor.  The zero eigenvalue is tangent to the line and changes the asymptotic baryon fraction.  The vacuum line
\begin{equation}
 D(u_0)=(0,0,u_0)
 \label{eq:vacuumline}
\end{equation}
has eigenvalues
\begin{equation}
 (-3,u_0-3,0).
 \label{eq:eigenD}
\end{equation}
For $0<u_0<3$ there are therefore two stable directions and one neutral direction tangent to the fixed-point line.  Each $D(u_0)$ possesses a two-dimensional local stable sheet consisting of trajectories with the same asymptotic value $u_\infty=u_0$.

\begin{figure*}[t]
\centering
\includegraphics[width=0.72\textwidth]{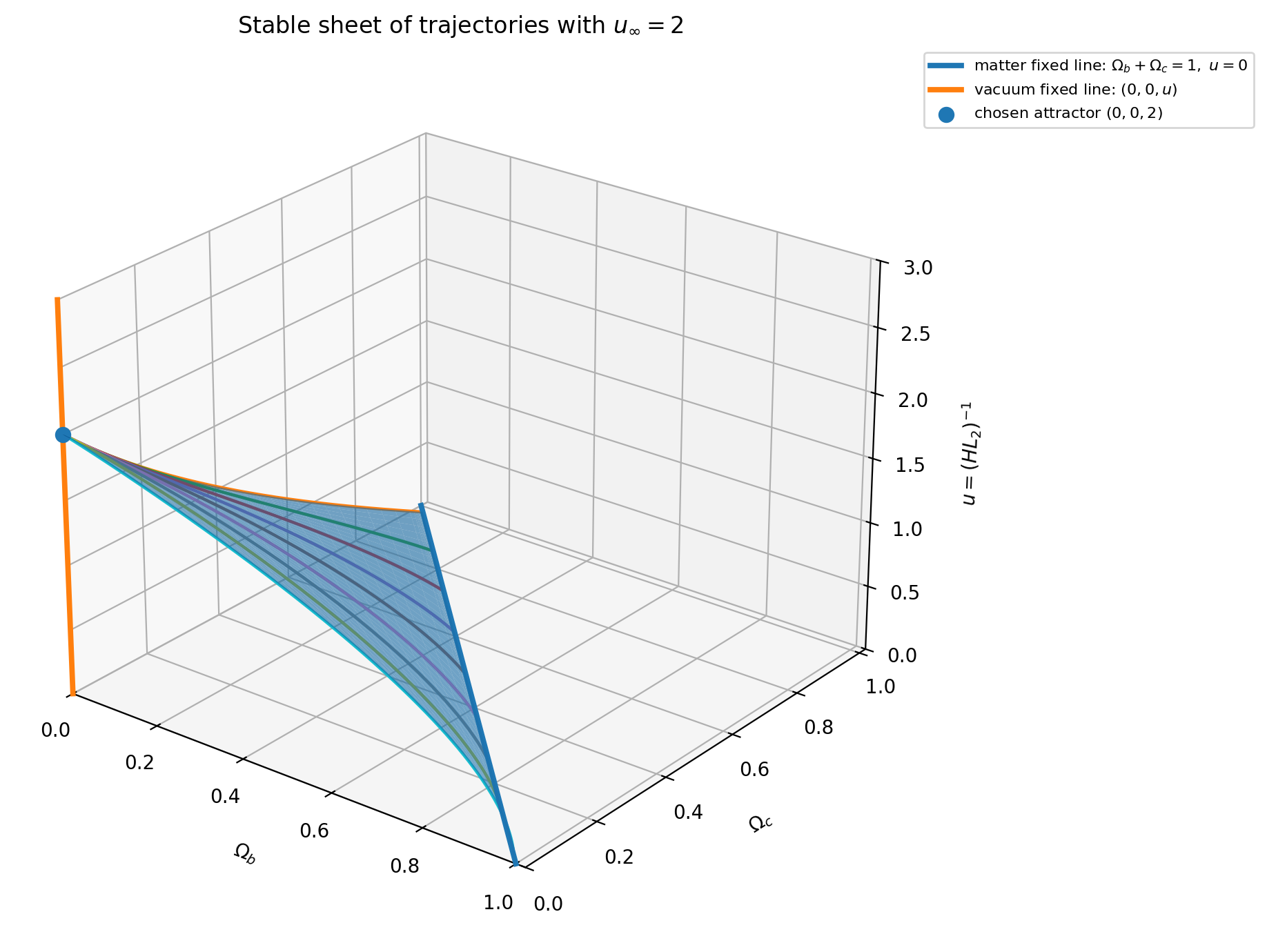}
\caption{Numerically constructed stable sheet ending at the representative attractor $D(2)=(0,0,2)$.  The sheet was obtained by forward shooting from a small-$u$ section near the matter line and selecting initial conditions with $u_\infty=2$.  Different curves on the sheet have different asymptotic baryon fractions in the past but share the same future endpoint.  The dark line on the base is the matter fixed-points line, while the vertical line is the vacuum fixed-points line.}
\label{fig:sheet}
\end{figure*}

For a mixed trajectory with $\Oc>0$ approaching $D(u_0)$, the late-time asymptotics are
\begin{align}
 \Ob&\sim B e^{-3N},\\
 \Oc&\sim C e^{-(3-u_0)N},\\
 u_0-u&\sim
 \frac{3u_0}{2(3-u_0)}\Oc,
 \qquad 0<u_0<3.
 \label{eq:3dasymptotic}
\end{align}
Thus CDM always decays at a slower rate than baryons near a finite-$u_0$ attractor.  Equivalently,
\begin{equation}
 \left(\ln\frac{\Oc}{\Ob}\right)'=u,
 \label{eq:ratio}
\end{equation}
so any nonzero CDM component eventually dominates the material sector when $u\to u_0>0$.

The endpoint $u_0=3$ is critical because the CDM eigenvalue in Eq.~\eqref{eq:eigenD} vanishes.  Linear stability theory is then insufficient, but the center dynamics is inherited from the universal model once baryons become negligible.  The leading approach is
\begin{equation}
 3-u\sim\frac{2}{N},
 \qquad
 \Oc\sim\frac{4}{9N^2},
 \qquad
 \Ob\propto e^{-3N}.
 \label{eq:critical3d}
\end{equation}
Thus the critical $u_0=3$ set asymptotically becomes the single-fluid separatrix of Sec.~\ref{sec:universal}: it is the natural phase-space continuation of the analytic thermogravity branch when a conserved baryon component is restored.

\subsection{The singular baryonic boundary}
\label{sec:baryonboundary}

The plane $\Oc=0$ is invariant, but its relation to the finite-$u_0$ sheets is subtle.  Setting $\Oc=0$ in Eqs.~\eqref{eq:Ob}--\eqref{eq:u3d} gives
\begin{equation}
 \frac{\dd\Ob}{\dd u}=\frac{2(\Ob-1)}{u},
\end{equation}
and hence
\begin{equation}
 \Ob=1+A u^2.
 \label{eq:baryonorbits}
\end{equation}
A baryonic trajectory that ends at a finite vacuum point $D(u_0)$ must therefore satisfy
\begin{equation}
\Ob=1-\frac{u^2}{u_0^2},\qquad \Oc=0.
 \label{eq:baryonLambda}
\end{equation}
Reconstructing the original variables gives
\begin{equation}
 H^2=\frac{\rho_b}{3}+\frac{1}{u_0^2L_2^2}.
 \label{eq:Lambdaeff}
\end{equation}
Thus the $\Oc=0$ edge of every finite-$u_0$ stable sheet is simply the conserved dust solution with a positive cosmological-constant-like integration constant,
\begin{equation}
 \Lambda_{\rm eff}=\frac{3}{u_0^2L_2^2}.
\end{equation}
This is the standard integration-constant freedom of trace-free gravity when stress energy is conserved~\cite{AndersonFinkelstein1971,HenneauxTeitelboim1989,NgVanDam1991,Smolin2009,PadillaSaltas2015}.

If instead that integration constant is set to zero, Eq.~\eqref{eq:Lambdaeff} gives
\begin{equation}
 \Ob=1,\qquad \Oc=0,
 \qquad u' = \frac32u,
 \label{eq:purebaryon}
\end{equation}
so the trajectory is a vertical line in $(\Ob,\Oc,u)$ and runs to $u\to\infty$: it does not terminate at any finite point on the vacuum line.  The limit $\Oc\to0^+$ is therefore non-uniform: for every arbitrarily small but nonzero CDM abundance Eq.~\eqref{eq:ratio} eventually amplifies the CDM-to-baryon ratio, whereas exactly $\Oc=0$ remains on the conserved baryonic plane forever.  This explains the apparent discontinuity at the edge of Fig.~\ref{fig:sheet}.  It is not a discontinuity of the vector field, but a singular late-time limit together with a choice of integration constant.

\section{Discussion and conclusions}
\label{sec:discussion}

The homogeneous thermogravity equations have a considerably richer global structure than is visible from the single analytic solution quoted in~\cite{IsicheiMagueijo2026,LeeEtAl2026}.  In the universal dust model the entire phase portrait is controlled by the first integral $c$.  The solution of Ref.~\cite{IsicheiMagueijo2026} is the critical orbit $c=-1/6$: it sits precisely between a family of asymptotically de Sitter trajectories and a family that turns around and recollapses.  This makes clear which part of the late-time acceleration is tied uniquely to the non-conservation scale $L_2$ and which part can be shifted by an additional integration constant. It would be interesting to examine the implications for the phenomenology discussed in \cite{LeeEtAl2026}.

The same distinction survives (in a geometrically more interesting form) when baryons are conserved and only CDM is sourced.  The one-dimensional vacuum attractor of the universal system becomes a line $D(u_0)$ in a three-dimensional phase space, and each normally attracting point $0<u_0<3$ is fed by a two-dimensional stable sheet.  The $u_0=3$ endpoint is critical and asymptotically CDM dominated, recovering the algebraic approach of the universal separatrix.  This is the branch most directly connected to the analytic solution used in Refs.~\cite{IsicheiMagueijo2026,LeeEtAl2026}.  If one restores covariant bookkeeping variables, the background energy transfer can be written in forms reminiscent of interacting dark-sector cosmologies~\cite{Wang2024,Zhai2025}, although the physical interpretation here is different: the exchange originates from a preferred thermodynamic frame rather than from a conventional dark-energy sector.

The conserved baryonic boundary also exposes an important conceptual point.  When $\Oc=0$, finite-$u_0$ endpoints require the usual cosmological-constant-like integration constant of trace-free gravity.  If one insists instead on the zero-integration-constant baryonic solution, the orbit remains at $\Ob=1$ and never reaches the finite-$u$ vacuum line.  Consequently, the operation of first taking the late-time limit and then sending the CDM abundance to zero does not agree with setting CDM to zero from the outset.  This singular limit should be kept distinct from the statement that vacuum energy itself does not gravitate in the underlying theory.

The observational CDM-only model of Ref.~\cite{LeeEtAl2026} includes radiation and is solved numerically over the full redshift range, whereas the present analysis deliberately isolates the late-time dust subsystem.  The phase-space picture nevertheless supplies a useful global map for those numerical solutions and identifies the critical surfaces and integration-constant freedoms that should be controlled when specifying cosmological boundary conditions.  Extending the analysis to radiation, spatial curvature and perturbations is straightforward in principle and will determine whether the critical structure identified here persists in the full cosmological system.

\section{Acknowledgments}
We would like to thank Dong Ha Lee 
for discussions that led to this paper. JM is partly supported by STFC Consolidated Grant ST/X000575/1 and Consolidated Award UKRI4230.
%ST/T000791/1.

\bibliographystyle{apsrev4-2}
\bibliography{references}

\end{document}